# Three-Dimensional Ultrasound Localization Microscopy Using Hadamard-Encoded Multipulse and 4D Denoising on a 256-Channel System with a 1024-element Matrix Array: A Feasibility Study in Human Kidney Transplants

**Authors' full names and affiliations:**

U-Wai Lok[1,†], Jingke Zhang[1,†], Jingyi Yin[1], Lijie Huang[1], Ryan M. DeRuiter[1], Yanzhe Zhao[1], Kaipeng Ji[1], Tao Wu[1], Qian Shi[2], Sanjeev Sethi[3], Kate M. Knoll[1], Dawn Boynton[1], Kathryn A. Robinson[1], Thomas D. Atwell[1], Andrew J. Bentall[4], Chengwu Huang[1,*], and Shigao Chen[1,*]

[†]Equally contributed to this work

[1]Department of Radiology, Mayo Clinic College of Medicine and Science, Rochester, MN 55905 USA

[2]Division of Clinical Trials and Biostatistics, Department of Quantitative Health Sciences, Mayo Clinic, Rochester, MN, USA

[3]Department of Laboratory Medicine and Pathology, Mayo Clinic, Rochester, MN, USA

[4]Division of Nephrology and Hypertension, Mayo Clinic, Rochester, MN 55905 USA

***Corresponding Authors: Shigao Chen, Ph.D. and Chengwu Huang, Ph.D**

Department of Radiology, Mayo Clinic, 200 First Street SW, Rochester, MN 55905, USA

Email: Chen.Shigao@mayo.edu and Huang.Chengwu@mayo.edu

## Abstract:

Three-dimensional ultrasound localization microscopy (3D ULM) enables volumetric visualization of microvasculature beyond the acoustic diffraction limit, but its clinical translation remains limited by long acquisition times, low microbubble (MB) signal-to-noise ratio (SNR), and insufficient MB contrast, particularly on ultrasound systems with a limited number of acquisition channels. In this study, we developed an in-human 3D ULM framework for a 256-channel research ultrasound system equipped with a 1024-element 32 × 32 matrix array. The framework combines diverging-wave Hadamard-encoded multipulse (HEM) transmission to enhance MB SNR with denoising strategy combining MB enhancement and four-dimensional non-local means filtering to improve MB contrast. The framework was implemented on a research ultrasound platform and evaluated in phantom experiments and four *in vivo* human kidney transplant studies. In phantom experiments, HEM improved the MB contrast ratio around 2 dB, while the proposed denoising strategy enhanced *in vivo* MB contrast by up to around 9 dB. For the 3D ULM reconstruction achieved a spatial resolution of approximately 84 μm, as estimated by Fourier shell correlation using the 1/2-bit criterion. The number of reconstructed microvascular voxels increased exponentially with acquisition duration, reaching approximately 74% of the estimated saturated vessel density using 3000 volumes (~12 s). Quantitative analysis within manually defined 3D renal cortical regions of interest enabled assessment of vessel density, flow velocity, and perfusion index. The subject with the highest histopathological injury exhibited reduced microvascular visibility and the lowest vessel density, flow velocities, and mean perfusion index. These results demonstrate that the proposed framework enables high-quality volumetric super-resolution imaging and quantitative characterization of renal microvasculature using a 256-channel ultrasound system, providing a practical step toward the clinical translation of *in vivo* 3D ULM.

# I. Introduction

Ultrasound localization microscopy (ULM) has emerged as a promising technique capable of visualizing microvascular structures at spatial scales well below the acoustic diffraction limit while retaining the penetration depth of conventional ultrasound [1-4]. ULM relies on the detection and localization of intravenously administered microbubbles (MBs). By accumulating the localized positions and trajectories of MBs over thousands of frames, ULM can reconstruct microvascular networks with micrometer-scale spatial resolution and simultaneously provide quantitative information on microvascular blood flow. The feasibility and potential utility of ULM have been demonstrated across a broad range of preclinical models and in clinical studies involving organs such as the brain, kidney, breast, and other tissues [4-7]. These advances have established ULM as a promising approach for characterizing microvascular morphology and hemodynamics *in vivo* [3, 4]. On the other hand, biological vascular networks, however, are intrinsically three-dimensional (3D). Two-dimensional (2D) ULM samples only a cross-sectional imaging plane and cannot recover the elevational positions of MBs or vessels outside that plane. Furthermore, MBs moving through tortuous 3D vessels may enter and leave the imaging plane, resulting in interrupted trajectories and potentially biased estimates of vascular morphology. Additionally, previous studies have shown that 2D ULM can underestimate velocity because microbubbles across the 3D flow profile are projected onto a single imaging plane [8]. These limitations have motivated the development of volumetric ULM. By localizing and tracking MBs in all three spatial dimensions, 3D ULM can recover the volumetric architecture of microvascular networks while avoiding the projection ambiguities inherent to 2D imaging.

Recent progress has been made toward practical 3D ULM. Early studies established volumetric super-resolution imaging using fully addressed 1024-element 32 × 32 matrix arrays and demonstrated that 3D ULM could resolve complex vascular structures beyond the diffraction limit [9, 10]. Subsequent studies [11-13] demonstrated 3D ULM using a matrix array with the same specifications, operated with a 256-channel ultrasound system. Additionally, high-volume-rate nonlinear imaging has further improved the robustness of 3D MB detection and tracking [14]. Furthermore, to improve the field of view, diverging cylindrical-wave transmission has been investigated to improve sensitivity and volume rate for multiplexed volumetric imaging [15]. On the other hand, A 256-element 16 × 16 matrix array has been developed [16] to reduce system complexity, but at the cost of a smaller aperture size and limited field-of-view coverage. Alternatively, to reduce the large channel count, different architectures have been explored, including sparse arrays [17] and row-column-addressed (RCA) arrays [18]. More recently, RCA-based implementations have enabled high-volume-rate 3D ULM with dramatically reduced channel requirements, including *in vivo* cerebral and renal applications [19-22].

Recent studies have also expanded the biological scope and computational capabilities of 3D ULM. Volumetric renal ULM has enabled visualization and quantitative assessment of renal microvascular structures, including glomeruli, and recent work has validated ULM-derived glomerular measurements against light-sheet microscopy [23]. Advances in beamforming sought to recover MB sensitivity and localization performance from matrix array acquisitions; for example, nonlinear delay-multiply-and-sum beamforming has been shown to improve MB tracking and vascular saturation relative to conventional delay-and-sum reconstruction [24]. At the same time, GPU-accelerated [25-28] strategies were used to address the substantial computational burden associated with volumetric datasets. A recent projection-based 3D ULM framework reformulated volumetric beamforming, clutter filtering, motion estimation, MB separation, and localization into computationally efficient lower-dimensional operations, achieving processing times approaching the acquisition duration in an *in vivo* porcine kidney model [29]. These advances highlight the rapid development of the 3D ULM field while also emphasizing the remaining barriers to clinical translation.

For clinical renal imaging, a sufficiently large field of view is needed to cover the kidney while maintaining the high volumetric imaging rate required for 3D ULM. Diverging wave transmission is therefore advantageous because it enables rapid volumetric imaging over an enlarged field of view. In addition, implementing 3D ULM on a 256-channel ultrasound system may provide a more cost-effective and practical approach for clinical translation than systems requiring simultaneous access to all 1024 elements of a matrix-array. However, performing *in vivo* 3D ULM with a conventional 1024-element 32 × 32 matrix array using diverging wave transmission on a 256-channel system presents several technical challenges. In particular, the limited number of simultaneously available channels requires multiplexing of the matrix-array elements, while diverging-wave imaging introduces trade-offs between field of view and MB quality (e.g. contrast and SNR). Addressing this challenge is therefore essential for developing clinically practical 3D ULM implementation.

This challenge is particularly important for in-human renal imaging. The kidney contains a dense and hierarchical vascular network extending from large renal vessels to cortical microvessels, making volumetric imaging desirable for assessing renal microvascular structure and function. Recent animal studies have demonstrated the potential of 3D ULM for renal microvascular imaging and quantitative analysis [30]; however, translating these capabilities to human kidneys is substantially more demanding because of increased imaging depth, acoustic attenuation, respiratory motion, and practical constraints on acquisition duration. Consequently, a 3D ULM framework capable of achieving sufficient MB sensitivity and vascular sampling using clinically practical hardware and acquisition times is needed to advance volumetric ULM toward human renal applications.

To address these challenges, we developed a 3D ULM acquisition and processing framework using a 256-channel ultrasound system equipped with a 1024-element 32 × 32 matrix-array and validated in human kidney transplants. The framework integrates several strategies designed to improve sensitivity in MB detection and reconstruction robustness. First, a diverging-wave Hadamard-encoded multipulse (HEM) transmission sequence was implemented to increase MB SNR, together with a customized GPU-based HEM decoding and diverging wave beamforming pipeline for efficient volumetric reconstruction. Second, an MB separation strategy was incorporated to improve the reliability of localization and tracking at increased MB concentrations, thereby improving vascular filling under the clinically feasible acquisition time. Third, we developed the use of MB enhancement with a four-dimensional (4D) non-local means (NLM) denoising method that jointly exploits three-dimensional spatial and temporal redundancies to enhance MB signals. The MB enhancement approach suppresses sidelobe artifacts and background noise and the 4D NLM can effectively suppresses background noise while preserving spatially and temporally coherent MB signals as compared with conventional lower-dimensional NLM approaches [31, 32], thereby improving MB detectability and localization for volumetric super-resolution imaging. Finally, 3D motion correction was performed to compensate for motion, followed by 3D MB localization and tracking to reconstruct the renal microvascular network and quantify blood-flow dynamics.

The performance of HEM was assessed in a phantom study. A vessel phantom was used to quantify the improvement in MB contrast achieved with the diverging wave HEM sequence. The complete framework was subsequently evaluated *in vivo* in human kidney transplants to assess MB contrast, microvessel spatial resolution, and vascular filling reconstruction. 3D regions of interests (ROI) were delineated on the reconstructed datasets to quantify microvascular parameters, including vessel density and flow velocity. Together, these experiments evaluate the feasibility of performing volumetric super-resolution imaging and quantitative microvascular assessment of the human kidney using a clinically practical 256-channel ultrasound platform. Four patients have been recruited and scanned, and quantitatively indices were compared with the histopathological biopsy results.

# II. Methods and Materials

*II-1. Hadamard-encoded multi-pulse diverging wave transmission*

In this study, Hadamard coding is applied to a sequence of $N$ diverging wave transmissions, each transmitted at a different steering angle [33]. A $N \times N$ Hadamard matrix is employed to modulate the polarity of each transmitted pulse across these angles. Each row of the matrix corresponds to a single transmission event and specifies the polarity configuration applied to the full set of angles. A matrix entry of +1 indicates that the transmitted pulse retains its positive polarity, while an entry of −1 represents a negative polarity. This polarity encoding strategy enables simultaneous transmission of multiple diverging waves, while maintaining the ability to recover angle-specific signals during the decoding process. As illustrated in Fig. 1(a), the encoding scheme is demonstrated using a $4\times4$ Hadamard matrix. In this case, four diverging wave transmissions are performed. The first transmission uses a uniform polarity pattern of all +1s, applying the same polarity across all steering angles. The fourth transmission applies an polarity pattern of [+1, −1, −1, +1]. Following an encoded transmission, the ultrasound transducer receives a superimposed radio-frequency (RF) signal containing the combined echoes from all $N$ consecutively transmitted wave pulses (Step 1 in Fig. 1b). Decoding is performed by applying the transpose of the Hadamard matrix to the vector of received signals. Because the Hadamard matrix is orthogonal, its transpose is equivalent to its inverse (with a scaling of $1/N$), enabling a straightforward and computationally efficient decoding process. Let $M$ denote the signals received for the Hadamard-coded transmission events. The decoding operation is performed by multiplying $M$ with the transpose of the Hadamard matrix $H^T$ as

$$S = H^T M, \tag{1}$$

where S is the resulting vector of decoded RF echo signals, each corresponding to a specific transmission angle. Theoretically, Hadamard encoding and decoding improves the signal-to-noise ratio (SNR) by a factor of square root of $N$. Additionally, after Hadamard decoding (as shown in step 2 in Fig. 1b), residual offsets remain among different transmissions (e.g., t1, t2, and t3 in step 3 of Fig. 1b). These offsets result from sequential transmissions with different steering angles. To maintain accurate spatial coherence during volumetric reconstruction, these transmission-dependent delays should be precisely compensated as shown in step 4 in Fig. 1(b).

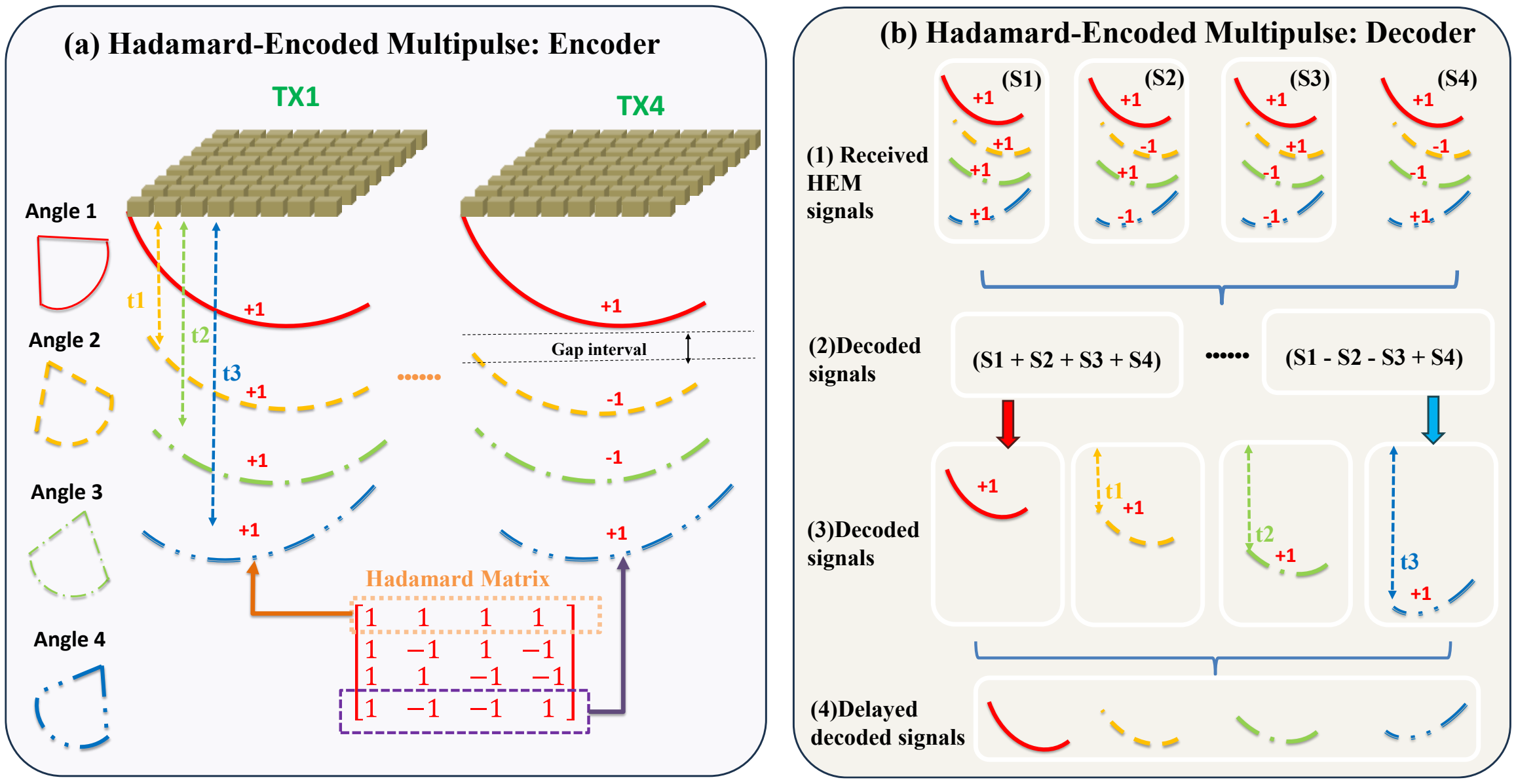

Fig. 1. Illustration diagram of the (a) encoder and (b) decoder process for Hadamard-encoded diverging wave imaging.

### *II-2. Vessel phantom study*

To evaluate the SNR improvement provided by the proposed method under the current imaging configuration, a vessel phantom experiment was performed. A 1024-element 32 × 32 matrix array transducer (Vermon, Tours, France), operating at a center frequency of 3.47 MHz, was used to image flowing microbubbles (MBs) in a customized vessel phantom (Computerized Imaging Reference Systems, Inc., Norfolk, VA, USA) containing a 2-mm-diameter vessel. Definity microbubbles (Lantheus, Bedford, MA) suspension was circulated through the vessel tube at a mean constant flow velocity of 3.87 mm/s. A total of 100 volumes were acquired for each imaging condition. To assess the benefit of Hadamard-encoded multipulse (HEM) transmission, MB signals were acquired without and with HEM sequentially with a time delay of around 2 seconds, first using HEM with fourth-order HEM and then without HEM. All acquired data were subsequently reconstructed both with and without HEM using post-processing.

*II-3. Monitoring of MB perfusion and data acquisition for in vivo study*

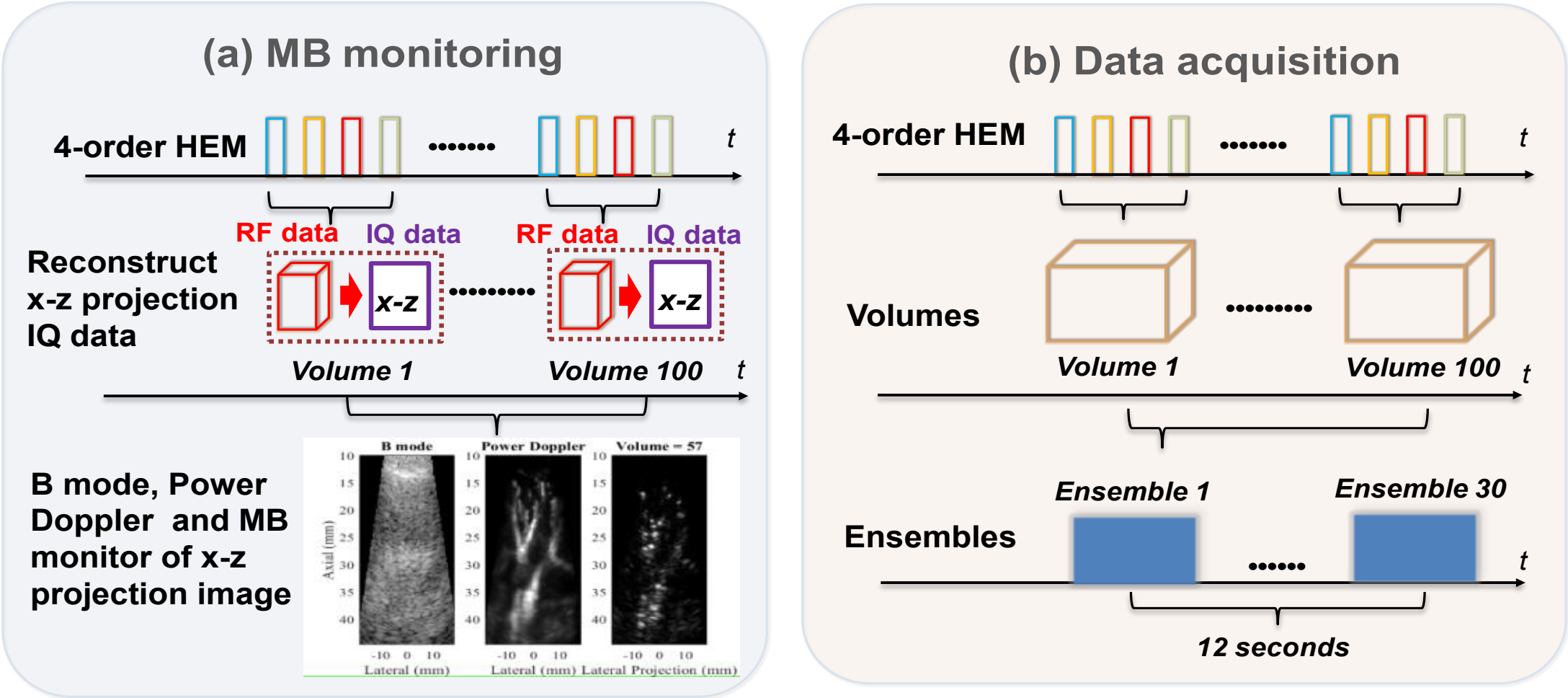

**Fig. 2.** (a) Schematic of MB monitoring approach for MB evaluation. (b) Schematic of RF channel data acquisition: each volume was acquired using four-angle HEM with diverging wave transmissions, and 100 volumes of RF data per ensemble were acquired using the Verasonics Vantage 256 system. The sequence was repeated 30 times to collect a total of 3000 volumes for post-processing.

The *in vivo* human study was approved by the Institutional Review Board (IRB) of Mayo Clinic, written informed consent was obtained from each participant. A Vantage 256 ultrasound system (Verasonics Inc., Kirkland, WA, USA) and a 1024-element 32 × 32 matrix array transducer (Vermon, Tours, France) were utilized for transplanted kidney data acquisition. The probe was connected to a Universal Transducer Adapter (UTA) 1024 MUX, which allows for the transmission of four sub-aperture (each sub-aperture has total of 32 × 8 elements) with different delays through multiplexing. The transmission strategy adopted a "light" transmission scheme as described in [34], wherein each sub-aperture receives signals not only from itself but also its adjacent apertures. This configuration requires 10 transmissions in total per HEM acquisition (total of 40 transmissions for 4-by-4 HEM encoding). The volunteers were instructed to hold their breath during each data acquisition session to minimize breathing tissue motion and then the RF channel data were acquired. As emphasized in [35], assessing the MB quality before ULM acquisition is critical to optimize imaging settings and timing of data acquisition. However, B-mode imaging alone was insufficient to assess vascular coverage and facilitate optimal probe positioning. Therefore, we developed an MB monitoring approach to facilitate the assessment of MB distribution, as shown in Fig. 2(a). In addition to a B-mode image reconstructed from the central imaging plane, a maximum-intensity projection of the power Doppler (PD) image onto the lateral–axial plane, with a projection thickness of approximately 10 mm, was displayed to facilitate identification of an appropriate probe position. Additionally, maximum-intensity projection images of the MB signals were displayed volume by volume to assess MB distribution and density to help determine the appropriate timing for channel data acquisition. The B-mode, projected PD, and projected MB images were available for display within 0.5 s after selecting the MB data monitoring mode (see the bottom panel of Fig. 2(a)), enabling semi-real-time assessment of temporal changes in MB distribution and density.

After preliminary MB evaluation, RF channel data were acquired as shown in Fig. 2(b). All patients were scanned using Definity (1 mL per bolus; Lantheus Medical Imaging, MA, USA) MBs. Data acquisitions were performed around 40 seconds after MB injection for each subject. A total of 3000 volumes (12 seconds) were collected to reconstruct the ULM volume, and the parameter settings are listed in Table I. Histological injury was graded using the Banff scoring [36] system based on biopsy samples obtained from the corresponding transplanted kidney within 48 hours after the ultrasound examination. In this study, four subjects were included to demonstrate the feasibility of quantitative evaluation of the injury of kidney transplants using 3D ULM images.

**Table I**. **Parameter settings for *in vivo* data acquisition**

| | Value |
|---|---|
| Transmit center frequency （MHz） | 3.47 |
| Sampling rate (MHz) | 13.9 |
| Pitch (mm) | 0.3 |
| Transmission type | Diverging wave 4-order HEM |
| Steering angles (Lateral, elevational) (degree) | (-5,0), (5,0), (0,-5), (0,-5) |
| Post-compounded volume rate (Hz) | 250 |
| Transmit one-sided voltage (V) | 10 |
| Number of pulse cycles after HEM decoding | 1 |
| Pulse repetition frequency (kHz) | 10 |
| Diverging angle (degree) | 15 |

### *II-4. Post-processing of the 3D ULM*

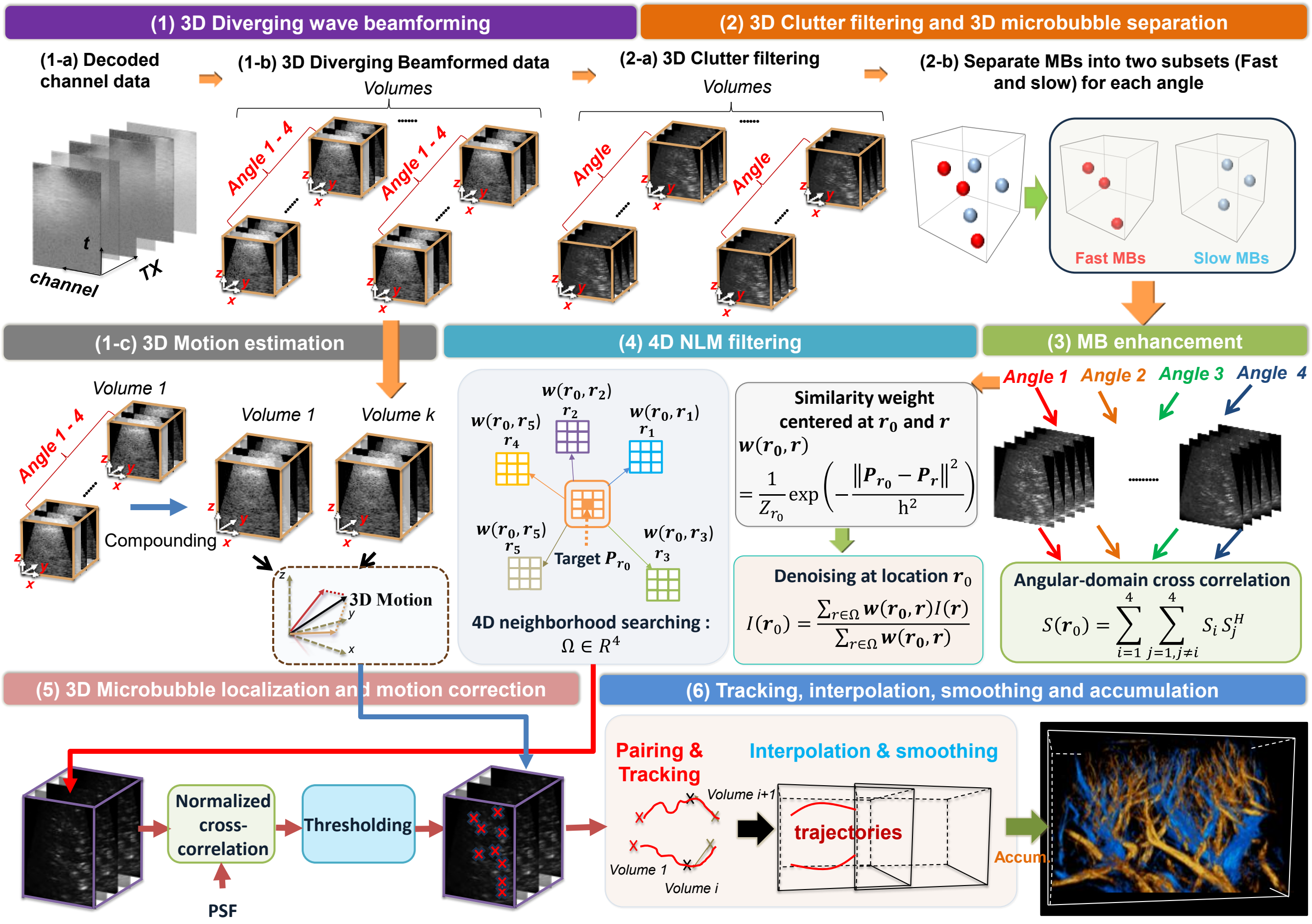

**Fig. 3.** Overview of the proposed 3D ULM processing framework. (1) 3D diverging-wave beamforming and motion estimation: (1-a) Hadamard-decoded channel data are beamformed to generate 3D volumes for each steering angle as shown in (1-b). (1-c) The beamformed volumes are compounded, and 3D tissue motion is estimated across consecutive volumes. (2) 3D clutter filtering and microbubble separation for each steering angle data: (2-a) 3D clutter filtering is applied to the beamformed data to suppress tissue signals and extract microbubble (MB) signals. (2-b) MBs are separated into fast- and slow-flow subsets for each steering angle. (3) MB enhancement: MB signals from different transmitting angles are enhanced using angular-domain cross-correlation. (4) 4D nonlocal-means (NLM) filtering: Spatiotemporal similarity within a 4D neighborhood is exploited to suppress noise while preserving MB signals. (5) 3D MB localization and motion correction:

MB locations are identified using normalized cross-correlation with the system point-spread function (PSF), followed by thresholding, and the estimated 3D tissue motion from 1-c is used to correct MB locations. (6) Tracking, interpolation, smoothing, and accumulation: Localized MBs are paired and tracked across consecutive volumes, followed by trajectory interpolation and smoothing. The resulting MB trajectories are accumulated to generate the final 3D super-resolution vascular image.

After HEM decoding, volumetric diverging-wave beamforming was performed to reconstruct the in-phase/quadrature (IQ) data with a pixel size of 0.5 λ (lateral) by 0.5 λ (elevational) by 0.5 λ (axial), where λ denotes the wavelength (around 0.44 mm). The volumetric data acquired at four steering angles were reconstructed separately and stored in memory for subsequent processing, as shown in step 1-b of Fig. 3. The compounded B-mode data were then used for motion estimation using a rigid, subpixel motion estimation algorithm. The resulting three-dimensional (3D) displacement estimates were stored for subsequent motion compensation after MB localization, as shown in step 1-c of Fig. 3.

Next, a 3D spatiotemporal singular value decomposition (SVD)-based clutter filter was applied to the pre-compounded IQ data (each transmitted angle) to separate moving MB signals from background tissue and stationary MB signals. The cutoff for the tissue clutter subspace was automatically determined using a lower-order singular-value thresholding method [37, 38]. Subsequently, an MB separation [39] method was applied to divide the original MB data into 4 subsets with sparser MB concentrations according to the speed and direction of MB motion.

Following MB separation, the proposed denoising method was applied in two stages: (1) MB enhancement and (2) four-dimensional non-local means (4D NLM) filtering, as shown in steps 3 and 4 of Fig. 3, respectively. The MB enhancement stage suppresses sidelobe artifacts and reduces the background noise level. MB cross-correlation similarity-based enhancement methods [40, 41] have been proposed to exploit the high correlation of signals originating from blood vessels or MBs from different transmissions while suppressing sidelobe artifacts and background noise, which generally exhibit lower correlation. Similar modified approaches have been successfully applied to both 2D and 3D microvascular imaging [42-46]. In conventional implementations, ultrasound signals are typically divided into two groups, and their cross correlation is evaluated to enhance coherent blood or MB signals while suppressing less correlated noise and artifacts. In the present study, rather than dividing the acquired MB signals into two groups, we exploited the correlation among MB signals obtained from different steering angles as illustrated in step 3 of Fig. 3, to suppress incoherent background components and sidelobe artifacts while preserving MB signals for subsequent processing. The resulting MB amplitude signals were subsequently processed using 4D NLM filtering to further suppress background noise and improve MB contrast [47].

After denoising, the MB signals were first normalized, and the MB signal envelopes were then thresholded to suppress background noise. Subsequently, 3D normalized cross-correlation (NCC) was performed between the MB data and a predefined point-spread function (PSF). The resulting NCC map was thresholded at 0.5 to remove low-correlation pixels, after which regional maxima were identified as MB centroids. Additionally, for each volume, motion correction was applied from the preceding motion-estimation step. MB trajectories were then constructed using a 3D bipartite graph-based pairing method [13]. Finally, the resulting MB trajectories were interpolated, smoothed, and accumulated to generate the 3D ULM image.

*II-5. Evaluation metrics for 3D ULM imaging*

To evaluate the spatial resolution of the resulting 3D ULM images, the vessel cross-sectional profiles were interpolated to a resolution 10 times higher than that of the original ULM profiles, and the full-width-half-maximum (FWHM) was subsequently measured. Additionally, Fourier shell correlation [48, 49] with the ½-bit threshold was used to obtain a global estimate of the 3D spatial resolution.

To compare the performance with and without HEM and denoising performance, the contrast ratio (CR) of MB signals was used to evaluate the improvement as

$$\mathrm{CR} = 10 \times \log_{10} \frac{MB_{mean}}{N_{mean}} , \quad (2)$$

where the $MB_{mean}$ is the mean intensity including microbubble signals and surrounding signals in the defined region-of-interest (ROI). $N_{mean}$ is the mean background intensity in the defined ROI.

Additionally, clinical ULM requires a sufficient number of localized microbubbles (MBs) within a short imaging period to enable vessel filling. The temporal evolution of vessel reconstruction illustrates the progressive development of the vascular network, transitioning from sparse and fragmented structures to a more continuous and saturated vascular map. A saturation curve was fitted using an exponential saturation model as

$$\mathrm{C(t)} = C_{eff}(1 - e^{-Kt}) \tag{3}$$

, where $\mathrm{C(t)}$ represents the number of vessel-containing voxels at acquisition time *t*, and $C_{eff}$ represents the extrapolated number of vessel-containing voxels at saturation within the entire field of view. *K* is a rate constant and the percentage of vessel filling was then defined as

$$\text{Percentage of vessel filling (\%)} = \frac{\mathrm{C(t)}}{C_{eff}} \times 100\% \tag{4}$$

. This metric represents the fraction of the extrapolated saturated vascular network recovered at a given acquisition time for the entire field of view.

*II-6. Quantitative metrics for 3D ULM*

Quantitative 3D ULM-derived metrics were computed within the manually segmented 3D cortical ROI to characterize renal microvascular morphology and hemodynamics. These metrics included vessel density (VD), defined as the percentage of voxels occupied by the reconstructed vascular network within the selected 3D ROI, as well as several flow velocity metrics, including mean flow velocity (MeanV), mean upward flow velocity, and mean downward flow velocity. Additionally, mean perfusion index (PI) was calculated as the product of vessel density and mean flow velocity (PI = VD × MeanV). Vessel density was expressed as a percentage (%), whereas all flow related metrics were reported in millimeters per second (mm/s). These quantitative indices were subsequently compared across four subjects to evaluate differences in renal microvascular perfusion. For each transplant subject, the biopsy derived pathology indices, including chronicity index and activity index were used as histopathological reference measures [50] to compare with the quantitative metrics derived from 3D ULM. The chronicity index was determined based on the Banff classification for kidney transplant pathology, including interstitial fibrosis (ci), tubular atrophy (ct), chronic glomerulopathy (cg), and vascular fibrous intimal thickening (cv). Each component was graded according to the Banff criteria, and the scores were combined (ci + ct + cv + 2cg) to provide an overall measure of chronic structural injury in the renal allograft. The activity index was assessed using Banff lesion scores reflecting active inflammatory and immune-mediated injury, including tubulitis (t), interstitial inflammation (i), glomerulitis (g), peritubular capillaritis (ptc), and intimal arteritis (v). Each component was graded according to the Banff criteria, and the scores were combined (t + i + v + g + ptc) to provide an overall measure of active allograft injury. Noted that complement component C4d was not included because it was not available in the pathology reports.

## III. Results

### *III-1. Contrast improvement with HEM and denoising*

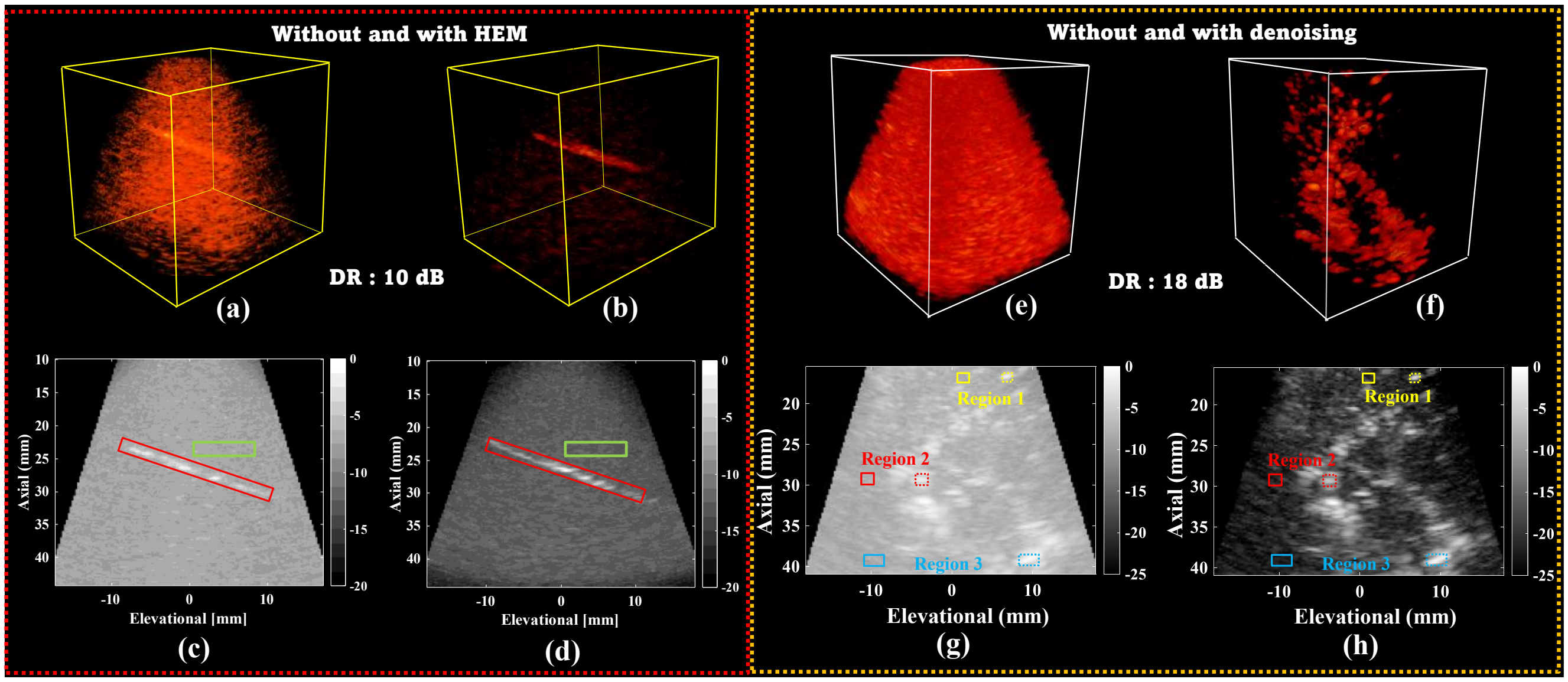

**Fig. 4.** 3D microbubble (MB) images (a) without and (b) with HEM. Both images are displayed using an identical dynamic range (DR) of 10 dB for direct comparison. A time delay of approximately 2 s occurred between the acquisitions with and without HEM. (c) Elevational-axial maximum-intensity projection (MIP) image without HEM and (d) corresponding MIP image with HEM. Red boxes indicate ROIs used to calculate the mean MB signal intensity, while green boxes indicate background ROIs used to calculate the mean background intensity. Both MIP images are displayed using an identical DR of 10 dB. 3D MB images (e) without and (f) with denoising. Both images are displayed using an identical DR of 18 dB for direct comparison. (g) Elevational-axial MIP image without denoising and (h) corresponding MIP image with denoising. Dashed boxes indicate ROIs used to calculate the mean MB signal intensity, while solid boxes indicate background ROIs used to calculate the mean background intensity.

Figures 4(a) and 4(b) show 3D renderings of MB volumes reconstructed without and with HEM respectively, using the vessel phantom with same dynamic range (10 dB). With HEM, background noise could be better suppressed and improves MB-to-background contrast. The corresponding elevational MIP images are shown in Figs. 4(c) and 4(d). The red and green boxes indicate the MB and background ROIs, respectively, used for CR assessment. With HEM, the CR increased from 0.61 dB (without HEM) to 2.64 dB (with HEM), further demonstrating improved MB-to-background contrast.

Figures 4(e) and 4(f) present 3D renderings of MB volumes reconstructed from volumetric ultrasound data without and with the denoising methods (MB enhancement and 4D NLM filtering) in the *in vivo* dataset, respectively. Both renderings are visualized with an identical dynamic range (18 dB) to enable direct comparison of noise suppression performance. As shown in Fig. 4(e), the MB volume reconstructed without denoising exhibits significant background noise, which hampers the detection and localization of individual MBs. In contrast, Fig. 4(f) demonstrates that applying the proposed denoising substantially suppresses background noise, resulting in higher MB-to-background contrast.

To further evaluate contrast enhancement, elevational MIP images are shown in Figs. 4(g) and 4(h) for the cases without and with denoising, respectively. Three ROIs, indicated by yellow (Region 1), red (Region 2), and blue (Region 3) boxes, were selected at varying depths for contrast ratio (CR) analysis. Dashed boxes correspond to vessel regions used to calculate the mean MB signal intensity, while solid boxes represent the adjacent background regions. The corresponding CR improvements are summarized in Table II. Compared with the case without denoising, the denoising improved the CR by 7.99 dB, 7.09 dB, and 8.95 dB in regions 1, 2, and 3, respectively.

**TABLE II. CRs with and without denoising.**

| | **CR (without denoising)** | **CR (with denoising)** |
|---|---|---|
| **Region 1** | 3.88 dB | 11.87 dB |
| **Region 2** | 4.09 dB | 11.18 dB |
| **Region 3** | 5.31 dB | 14.26 dB |

*III-2. Evaluation of 3D ULM images*

Figure 5(a) presents the elevational–axial maximum intensity projection (MIP) image of a kidney transplant with a low histological injury (all the Banff score = 0) obtained using the proposed method. To quantitatively evaluate the spatial resolution, the FWHMs of two microvessels were measured (as shown in yellow-dotted line in Fig. 5(a) to be approximately 89 µm and 86 µm, as shown in Fig. 5(b). The corresponding 3D bidirectional super-resolution images are shown in Fig. 5(c). A magnified view is presented in Fig. 5(d), as indicated by the white arrows, 3D microvascular structures near the renal cortex can be clearly visualized, demonstrating the capability of the proposed framework to resolve the cortical microvasculature *in vivo*. Additionally, the 3D upward and downward flows can be visualized separately, as shown in Fig. 5(e), enabling direction-specific quantitative analysis of both vascular morphology and flow velocity. The Fourier shell correlation (FSC) method with the ½-bit threshold criterion was applied to estimate the global 3D spatial resolution. As shown in Fig. 5(f), the estimated resolution was around 84.8 µm.

In addition, 3D velocity-resolved microvascular imaging was performed to further characterize renal hemodynamics, as shown in Fig. 5(g). The 3D velocity map, representing flow velocity magnitude, visualizes the spatial distribution and flow dynamics of microbubbles within the vascular network, enabling simultaneous assessment of vascular architecture and blood flow characteristics. The corresponding velocity histogram is presented in Fig. 5(h), illustrating the distribution of blood flow velocities across the reconstructed microvasculature, with upward and downward flow represented by the positive and negative velocity ranges, respectively. An exponential saturation model [Eq. (3)] was fitted to characterize the relationship between the number of acquired volumes (acquisition time) and vascular network recovery, as indicated by the white dotted curve in Fig. 5(i). Under the current acquisition protocol, approximately 74% of the extrapolated saturated number of vessel-containing voxels [Eq. (4)] was recovered within the 12 s acquisition.

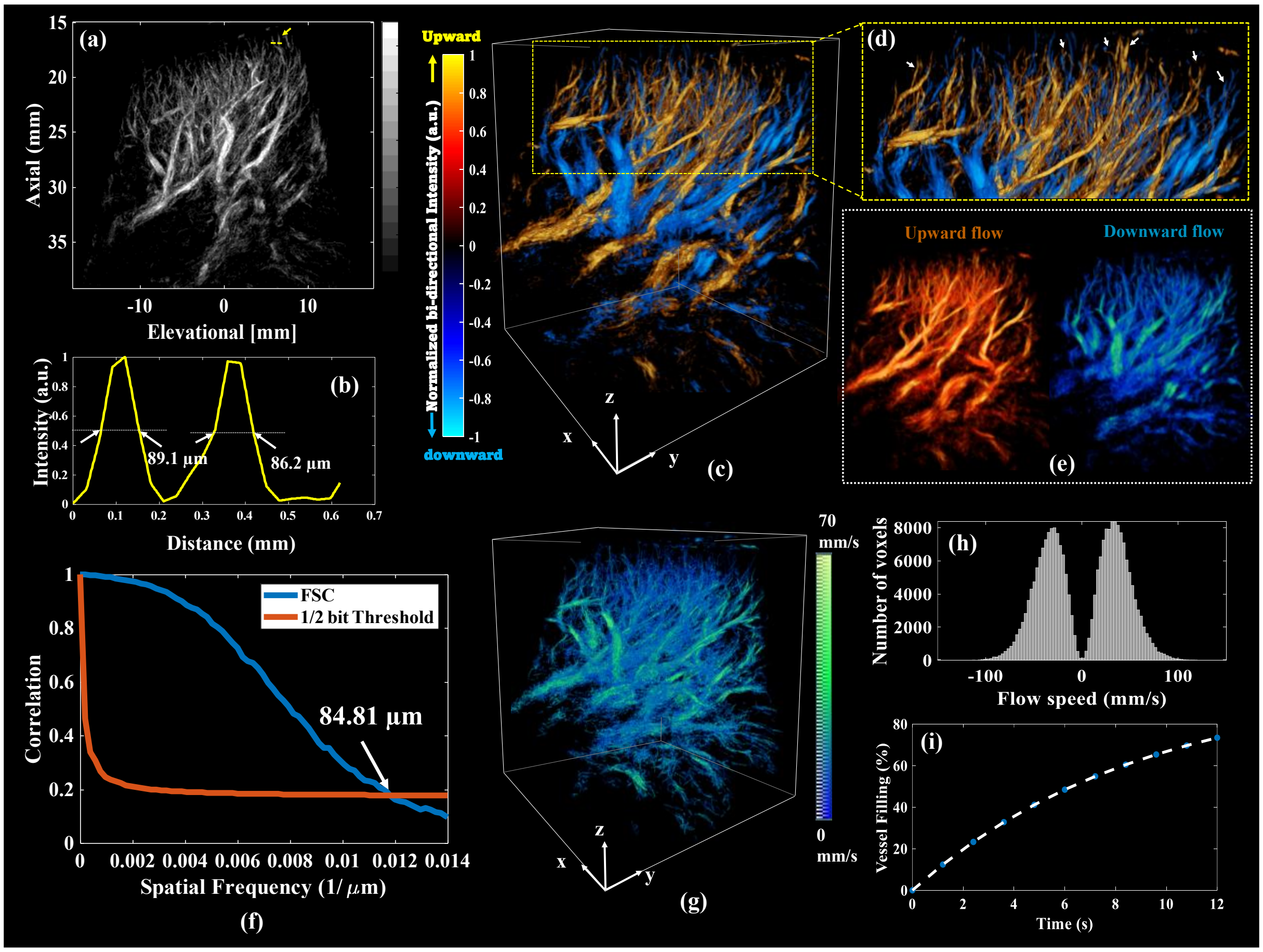

Fig. 5. (a) Elevational–axial MIP image of the reconstructed 3D ULM image from an *in vivo* kidney transplant. (b) Corresponding cross-sectional intensity profiles along the yellow dashed lines in (a), demonstrating that blood vessels as small as 89 µm and 86 µm can be resolved. (c) Bidirectional flow map of the reconstructed 3D ULM volume. (d) Magnified view of the region outlined by the yellow dashed box in (d) for the 3D ULM image. The white arrows highlight representative

microvessels, demonstrating that the proposed method enables visualization of 3D microvascular structures. (e) Upward and downward flow of the 3D ULM, (f) Fourier shell correlation curve used to estimate the spatial resolution of the reconstructed 3D ULM volume. The spatial resolution is determined from the intersection of the FSC curve with the 1/2-bit threshold criterion. (g) Three-dimensional flow velocity map reconstructed from the ULM data. (h) Distribution of flow velocities, separated into upward and downward flow components. (i) Exponential saturation fitting of the number of voxels as a function of time. Approximately 74% of the extrapolated saturated number of vessel-containing voxels is achieved using 3000 acquired volumes (12 seconds).

*III-3. Quantitative 3D ULM*

To enable quantitative analysis across subjects, 3D ROIs were manually delineated within the renal cortex for all 3D ULM datasets, as indicated by the white regions in Fig. 6(a–d). Specifically, the cortical ROI was delineated slice by slice in 2D and subsequently interpolated to generate a 3D mask. Within each ROI, quantitative microvascular metrics, including vessel density, mean flow speed, mean upward and downward flow speeds, and mean perfusion index, were calculated to characterize the renal microvascular network. The chronicity and activity indices were used to characterize the degree of histopathological injury, with higher values generally indicating greater injury. Among the four subjects, Subject 4 exhibited the highest chronicity and activity indices, suggesting the greatest degree of histopathological injury. Qualitative assessment of the corresponding super-resolution ULM images showed reduced microvascular visibility in Subject 4 compared with Subjects 1–3 [Fig. 6(d)]. In addition, fewer microvessels exhibiting upward flow were visually observed in Subjects 3 and 4 than in Subjects 1 and 2.

The corresponding quantitative ULM measurements are summarized in Table III. Consistent with the qualitative observations, Subject 4 exhibited the lowest vessel density (21.32%), whereas Subjects 1–3 showed relatively higher vessel densities, ranging from 29.41% to 34.66%. Subject 4 also exhibited the lowest mean flow speed and mean upward flow speed. In contrast, Subject 2 demonstrated higher mean upward and downward flow speeds (52.53 and 49.21 mm/s, respectively) and a higher mean perfusion index (17.47 mm/s) than the other subjects. Furthermore, the mean perfusion index, which incorporates both vessel density and mean flow speed, was lower in Subjects 3 and 4 than in Subjects 1 and 2. Collectively, these findings suggest reduced microvascular perfusion in Subjects 3 and 4, particularly in Subject 4, which also exhibited the greatest histopathological injury.

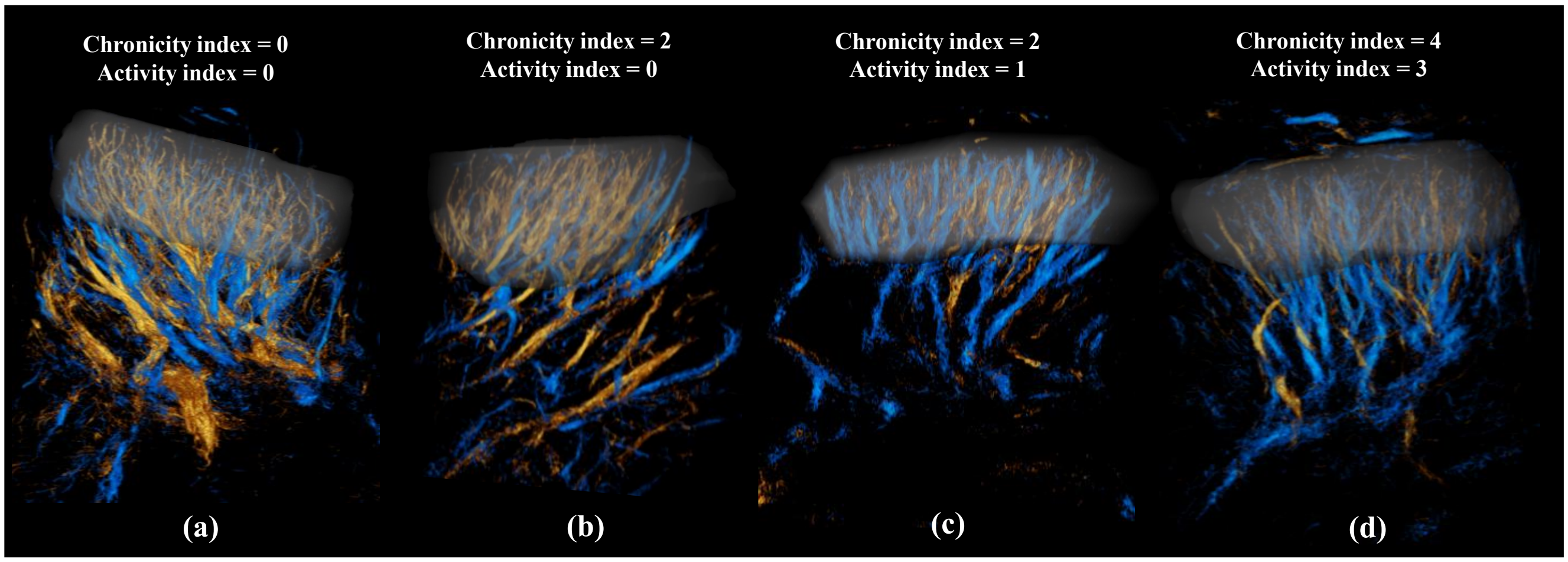

Fig. 6. Representative 3D ULM images from different subjects. The white regions indicate the selected regions of interest (ROIs) used for quantitative analysis.

**TABLE III. 3D ULM quantification indices for kidney allografts.**

| | Vessel Density (%) | Mean flow speed (mm/s) | Mean downward flow (mm/s) | Mean upward flow (mm/s) | Mean perfusion index (mm/s) | Chronicity index | Activity index |
|---|---|---|---|---|---|---|---|
| **Subject 1** | 34.66 | 40.10 | 39.69 | 40.35 | 13.90 | 0 | 0 |
| **Subject 2** | 33.62 | 51.81 | 49.21 | 52.53 | 17.42 | 2 | 0 |
| **Subject 3** | 29.41 | 39.52 | 38.96 | 39.84 | 11.62 | 2 | 1 |
| **Subject 4** | 21.32 | 39.18 | 39.98 | 39.02 | 8.35 | 4 | 3 |

## IV. Discussion

In this study, we demonstrated the feasibility of 3D ULM for high-resolution visualization and quantitative characterization of the renal microvasculature using a 256-channel ultrasound imaging system equipped with a 1024-element 32 × 32 matrix array. To improve MB signal quality during volumetric acquisition, a diverging-wave Hadamard encoding and decoding (HEM) strategy was implemented. The use of diverging waves is to provide a large field of view while Hadamard encoding increased the effectively transmitted acoustic energy while maintaining a reasonable volumetric rate required for tracking rapidly moving MBs, resulting in an improved MB SNR. Next, the proposed denoising strategy was further employed to suppress background noise. The combination of enhanced transmission and post-processing-based noise suppression therefore improves MB detectability and provides more reliable signals for subsequent localization and tracking. The proposed framework achieved a global spatial resolution of 84.8 μm, as estimated using Fourier shell correlation (FSC) with the 1/2-bit threshold criterion. This resolution substantially exceeds the theoretical diffraction-limited resolution of approximately 443 μm, corresponding to one acoustic wavelength under the imaging conditions used in this study. Furthermore, approximately 74% of the extrapolated saturated number of vessel-containing voxels was recovered within the 12 s acquisition. These results suggest that a reasonable portion of the renal microvascular network can be reconstructed within a relatively short acquisition period.

Beyond morphological visualization of the microvasculature, the proposed 3D ULM framework enables quantitative assessment of blood flow velocity and direction, providing functional information that complements vascular morphology. For quantitative microvascular assessment, 3D ROIs were manually delineated within the renal cortex, from which vessel density, flow velocity, and perfusion index were quantified, as summarized in Table III. Notably, the subject with the highest degree of histopathological injury exhibited reduced microvascular visibility and consistently lower quantitative measures, including the lowest vessel density, mean flow velocity, and mean perfusion index, compared with the other subjects. These findings demonstrate the feasibility of the proposed framework for simultaneous volumetric visualization and quantitative characterization of renal microvascular morphology and hemodynamics *in vivo* using a 256-channel ultrasound system.

Several opportunities for further optimization for both system and algorithms should be noted. First, computational efficiency remains an important consideration for the implementation of 3D ULM. Volumetric beamforming, clutter filtering, and ULM reconstruction involve substantially larger datasets and may require several hours of CPU-based processing, although GPU acceleration can considerably reduce the processing time. Continued advances in computational efficient algorithm design are expected to further improve processing efficiency. For example, a recent study [29] reformulated major components of the 3D ULM processing pipeline, including clutter filtering, motion estimation, and MB localization, as a series of computationally efficient 2D operations while maintaining comparable 3D reconstruction performance using a row-column array. Incorporating similar computational strategies into the proposed framework may further accelerate data processing and represent a direction toward near-real-time 3D ULM reconstruction.

The implementation of the 4D NLM filter also provides opportunities for further optimization. In this study, 100 volumetric datasets were processed within each temporal window to accommodate the available GPU memory while maintaining practical computational requirements. A longer temporal window could potentially provide additional information for denoising, although its benefit relative to the associated computational and memory requirements remains to be systematically evaluated. More memory-efficient GPU implementations, parallel computing, or distributed processing could enable larger temporal windows and facilitate further investigation of this tradeoff. In addition, the filtering parameter $h$ (as shown in step-4 in Fig. 3) was empirically selected in the current implementation. Automated selection strategies based on data characteristics or noise statistics could further standardize this process and improve the adaptability of the framework across different imaging conditions.

The 2D matrix array used in this study, combined with the proposed Hadamard-encoded diverging-wave transmission strategy, improved MB sensitivity. Further improvement in transmit efficiency could extend the applicability of the framework to deeper anatomical targets and a broader range of subjects. Advanced transmission approaches, including other coded excitation [51, 52], and optimized receive amplification [53, 54], could potentially increase the effective SNR and improve MB detectability at greater imaging depths.

In this study, 3D ROIs were manually selected to ensure anatomically appropriate analysis. Previous studies have demonstrated the feasibility of kidney segmentation directly from 3D ultrasound using statistical shape models [55, 56], more recently, deep-learning methods [57, 58]. Future work could therefore investigate segmentation strategies specifically optimized for volumetric ULM.

Importantly, the current imaging protocol acquired 12 seconds within a single breath-hold, and the implemented 3D motion estimation and correction approach was sufficient to compensate for tissue displacement during this acquisition period. This acquisition strategy provides a practical means of minimizing respiratory motion while preserving sufficient MB observations for microvascular reconstruction. For applications requiring longer acquisition periods or free-breathing imaging, more advanced non-rigid motion compensation could further improve robustness to physiological motion. Recent learning-based registration approaches, such as VoxelMorph-based models [59] incorporating bidirectional similarity constraints and attention mechanisms, provide computationally efficient estimation of dense 3D displacement fields without iterative optimization or ground-truth motion labels.

Finally, this feasibility study included four subjects for evaluation of ULM-derived quantitative metrics. Despite the relatively small cohort, the present results establish the feasibility of using the proposed 3D ULM framework to obtain volumetric microvascular and hemodynamic measurements *in vivo* and provide a foundation for subsequent validation studies. Studies involving larger and more diverse cohorts will be valuable for characterizing intersubject variability, evaluating the reproducibility of ULM-derived metrics, and determining their associations with histopathological findings. Such investigations will be important for establishing the potential of 3D ULM-derived microvascular parameters as noninvasive imaging biomarkers for kidney allograft assessment.

## Conclusion

The proposed 3D ULM framework enables volumetric super-resolution microvascular imaging using a 256-channel ultrasound system with a 1024-element 32 × 32 matrix array. By integrating Hadamard-encoded diverging wave imaging with advanced denoising and 3D microbubble localization and tracking, the framework addresses key technical challenges associated with *in vivo* 3D ULM. Using this framework, 3D renal microvascular morphology, as well as blood flow velocity and direction, can be visualized and quantified. Furthermore, we demonstrate the feasibility of quantitatively assessing transplanted kidneys using 3D ULM-derived microvascular metrics in association with histopathological biopsy findings. Collectively, these results represent an important step toward the clinical translation of *in vivo* 3D ULM as a noninvasive approach for assessing microvascular structure and hemodynamics in kidney transplants.

**Acknowledgement**

This project was supported partially by the National Institute of Diabetes and Digestive and Kidney Diseases under award numbers R01DK138998 and R01DK129205. The content is solely the responsibility of the authors and does not necessarily represent the official views of the NIH. The Mayo Clinic and some of the authors (J.Z., C.H. and S.C.) have a potential financial interest related to the technology referenced in the research.